\documentclass[apjl]{emulateapj}
\usepackage{graphicx}

\def\etal{{et al.~}}

\begin{document}

\title{The Cosmic Evolution of AGN in Galaxy Clusters}

\author{Audrey Galametz\altaffilmark{1,2,3}, Daniel Stern\altaffilmark{1},
Peter R. M. Eisenhardt\altaffilmark{1}, Mark Brodwin\altaffilmark{4},
\\ Michael J. I. Brown\altaffilmark{5}, Arjun Dey\altaffilmark{4},
Anthony H. Gonzalez\altaffilmark{6}, Buell T. Jannuzi\altaffilmark{4},
\\ Leonidas A. Moustakas\altaffilmark{1}, S. Adam Stanford\altaffilmark{7}}

\altaffiltext{1}{Jet Propulsion Laboratory, California Institute of Technology, 4800 Oak Grove Dr., Pasadena, CA 91109, [e-mail: {\tt agalamet@eso.org}]}
\altaffiltext{2}{European Southern Observatory, Karl-Schwarzschild-Str. 2, D-85748 Garching, Germany}
\altaffiltext{3}{Observatoire Astronomique de Strasbourg, 11 rue de l$'$Universit\'e, 67000 Strasbourg, France}
\altaffiltext{4}{National Optical Astronomy Observatory, 950 North Cherry Avenue, Tucson, AZ 85719}
\altaffiltext{5}{School of Physics, Monash University, Clayton, Victoria 3800, Australia}
\altaffiltext{6}{Department of Astronomy, University of Florida, Gainesville, FL 32611}
\altaffiltext{7}{Institute of Geophysics and Planetary Physics, Lawrence Livermore National Laboratory, Livermore, CA 94550}

\begin{abstract} 

We present the surface density of luminous active galactic nuclei
(AGN) associated with a uniformly selected galaxy cluster sample
identified in the $8.5$ square degree Bo\"{o}tes field of the NOAO
Deep Wide-Field Survey.  The clusters are distributed over a large
range of redshift ($0<z<1.5$) and we identify AGN using three
different selection criteria:  mid-IR color, radio luminosity, and
X-ray luminosity.  Relative to the field, we note a clear overdensity
of the number of AGN within $0.5$ Mpc of the cluster centers at $z > 0.5$.  
The amplitude of this AGN overdensity increases with redshift.  Although there
are significant differences between the AGN populations probed by
each selection technique, the rise in cluster AGN surface density
generally increases more steeply than that of field quasars.  In
particular, X-ray selected AGN are at least three times more prevalent
in clusters at $1 < z < 1.5$ compared to clusters at $0.5 < z < 1$.
This effect is stronger than can be explained by the evolving median
richness of our cluster sample.  We thus confirm the existence of
a Butcher-Oemler type effect for AGN in galaxy clusters, with the
number of AGN in clusters increasing with redshift.

\end{abstract}


\section{Introduction}

In the local universe, AGN are known to be much less common in rich
galaxy cluster environments as compared to the field.  Optical
observations dating back more than three decades show that while
5\% of massive field galaxies show spectroscopic signatures of
nuclear activity, only 1\% of the corresponding population in local
galaxy clusters show such signatures \citep{Dressler1985}.  Assuming
that the $M_\bullet-\sigma$ relation \citep[e.g.,][]{Magorrian1998} holds
in rich environments, this suggests that the supermassive black
holes hosted by massive cluster galaxies are currently quiescent
or their nuclear activity is optically obscured.  In addition, the
intracluster plasma in rich galaxy clusters  is maintained at high
temperatures; AGN feedback has been suggested as the most natural
heat source by numerous authors \citep[e.g.,][]{McNamara2000,Birzan2004}.
This suggests that the activity level of AGN in cluster galaxies,
and their influence on the formation and evolution of clusters, may
have been higher in the past.

Early studies of AGN in clusters largely relied on optical spectroscopy
\citep[e.g.,][]{Bahcall1969}, a technique which is notoriously
insensitive to both obscured and low luminosity AGN.  In contrast,
multi-wavelength approaches provide a significantly more thorough
census of AGN.  For example, Compton-thick, luminous AGN will be
faint at the relatively low ($< 10$~keV) energies probed by {\it
Chandra} and {\it XMM/Newton} and would be unlikely to show a
UV-excess or spectroscopic signatures of nuclear activity at
cosmological distances.  Such sources would be easier to identify
at radio or mid-IR energies, though only a $\sim 10\%$ of AGN are
radio-loud.  We briefly review selection of AGN at non-optical
wavelengths, summarizing previous work using these techniques to
study the cosmic evolution of AGN in galaxy clusters.

{\it X-Ray Selected AGN.}  In the past decade, sensitive X-ray
observatories have provided an efficient method to identify AGN.
Several studies conducted on individual clusters or small cluster
samples have shown an excess of X-ray point sources in galaxy
clusters \citep{Cappi2001, Eckart2006}.  Martini \etal (2002, 2006,
2007)\nocite{Martini2002, Martini2006, Martini2007} study samples
of X-ray sources near low-redshift galaxy clusters with spectroscopic
follow-up to confirm cluster membership.  Similar to field X-ray
source studies, this work has shown that X-ray emitting galaxies
in clusters often do not show AGN signatures in their optical
spectra, implying that a significant fraction of cluster nuclear
activity is optically obscured.  Considering a sample of four galaxy
clusters at $z \sim 0.6$, \citet{Eastman2007} show that the X-ray
AGN are more common at $z \sim 0.6$ than at $z \sim 0.2$, implying
that the cluster AGN fraction increases rapidly with redshift.

Other groups have foregone spectroscopic observations and have
looked for enhancements in the radial distribution of X-ray sources
around clusters.  For example, \citet{Ruderman2005} derive the
surface density profile of X-ray sources around a sample of $51$
galaxy clusters at $0.3 < z < 0.7$ from the {\em Chandra} MAssive
Cluster Survey \citep[MACS;][]{Ebeling2001}.  They find that X-ray
sources show a very pronounced central spike ($r < 0.5$ Mpc) and a
second excess at $\sim 2.5$ Mpc relative to the cluster centers.
They suggest the former is from nuclear activity triggered by
infalling galaxies approaching the central giant elliptical galaxy,
while the second peak, which is most pronounced in virialized
clusters, is due to galaxy mergers at the cluster-field interface.
\citet{Bignamini2008} find an excess of X-ray point sources near
the centers of optically-selected clusters at $0.6 < z < 1.2$.

{\it Radio Selected AGN.}  Over the past $50$ years there have been
numerous studies of the environments of powerful radio galaxies
which are now well known to often be associated with the central
member of galaxy clusters, not only at low redshifts \citep{Matthews1964}
but also at high redshifts \citep[][Galametz et al., in
prep.]{Dickinson1993, Stern2003, Miley2004, Venemans2005, Venemans2007}.
Correlating the Sloan Digital Sky Survey \citep[SDSS;][]{Gunn1995}
with the Faint Images of the Radio Sky at Twenty cm survey
\citep[FIRST;][]{Becker1995}, \citet{Best2005} show that there is
a strong tendency for low-redshift radio-loud galaxies to be found
in rich environments. Lin \& Mohr (2007)\nocite{Lin2007} derive the
radio-loud AGN density profile from a sample of $573$ nearby ($z
\leq 0.2$) X-ray selected galaxy clusters and find a clear excess
of AGN in the center of the galaxy clusters.  \citet{Croft2007},
studying brightest clusters galaxies (BCGs) at $z < 0.3$ from a
sample of $13,240$ galaxy clusters, report that $19.7$\% of BCGs
are detected in the radio, with the majority of these ($92.8$\%)
being radio-loud ($L_{\rm 1.4GHz} \ge 10^{23}$ W Hz$^{-1}$).  In
contrast to these studies of X-ray and radio-selected AGN in galaxy
clusters, optically-selected, luminous quasars exhibit clustering
comparable to that of local field galaxies, and therefore primarily
reside in lower mass dark matter halos \citep[e.g.,][]{Croom2005,
Coil2007}.

{\it Mid-IR Selected AGN.}  AGN can also be efficiently identified
using mid-infrared colors \citep{Lacy2004, Stern2005}.  While the
optical to mid-infrared ($\lambda \leq 5 \mu$m) continuum of normal
galaxies is dominated by the composite stellar black body emission
and peaks at approximately $1.6 \mu$m, the optical to mid-infrared
continuum of active galaxies is dominated by a power law.  Consequently,
with sufficient wavelength baseline, one can easily distinguish AGN
from stellar populations.  Little work has been done to date on the
prevalence of mid-IR selected AGN in clusters, though Hickox et al.
(submitted) shows that low redshift ($z < 0.8$) mid-IR selected
AGN have weaker clustering properties (e.g., smaller correlation lengths) than either X-ray or
radio-selected AGN. At high redshift ($z \sim 2$), however, various
mid-infrared  AGN samples show strong clustering, suggesting that
they are preferentially associated with clusters \citep{Brodwin2008,
Dey2008, Magliocchetti2008}.

Previous studies of AGN activity in galaxy clusters have typically focussed
on a single selection technique, generally over a limited
redshift range and with a limited cluster sample size.  Combining
the IRAC Shallow Survey \citep[ISS;][]{Eisenhardt2004} and the
{\it Chandra} XBo\"{o}tes Survey \citep{Murray2005}, \citet{Gorjian2008}
compare mid-IR and X-ray AGN selection and find that only $28$\%
of XBo\"{o}tes sources are selected by the \citet{Stern2005}
mid-infrared criteria. On the other hand, a significant 
number (43\%) of AGN selected in mid-infrared have an X-ray counterpart
in XBo\"{o}tes (V.Gorjian, private communication).  Furthermore,
only $\sim 10$\% of optically-selected luminous quasars are radio-loud
\citep[e.g.,][]{Stern2000}.  These results imply that to obtain the
fullest census of nuclear activity, multiple AGN selection techniques
are required.

In this paper, we present the surface density of AGN
selected by three techniques (mid-infrared color, radio luminosity,
and X-ray luminosity) in a large sample of uniformly-selected galaxy clusters
identified in the Bo\"{o}tes field \citep{Eisenhardt2008}.  The
next section (\S 2) describes the galaxy cluster and AGN samples.
We describe the radial surface density profile of AGN in galaxy
clusters in \S 3 and discuss the results in \S 4.  We assume a
$\Lambda$CDM cosmology with $H_0 = 70$ km s$^{-1}$ Mpc$^{-1}$,
$\Omega_m = 0.3$ and $\Omega_{\Lambda} = 0.7$.

\section{The Galaxy Cluster and AGN Samples}

The galaxy cluster sample we analyze has been uniformly selected
from the 8.5 square degree ISS field in Bo\"{o}tes survey, which has been observed
by a wide array of facilities at multiple wavelengths. The following
sections describe first the cluster sample and then AGN samples
identified in the Bo\"{o}tes field at mid-infrared, radio and X-ray
energies. Note that in the later analysis (\S 3), we will use these
AGN samples, generally from flux-limited catalogs, to study the
cosmic evolution of AGN in galaxy clusters. To ensure uniformity
in the analysis, this will entail creating luminosity-selected
sub-samples (\S 3).

\subsection{The IRAC Shallow Survey cluster sample}

\citet{Eisenhardt2008} present a sample of $335$ galaxy clusters
identified in the Bo\"{o}tes field of the NOAO Deep Wide-Field
Survey \citep[NDWFS;][]{Januzzi1999}. The clusters, reaching out
to redshift $z = 2$, were identified by combining ISS with $B_WRI$
imaging from the NDWFS and $JK_s$ imaging from the FLAMINGOS
Extragalactic Survey \citep[FLAMEX;][]{Elston2006}.  \citet{Brodwin2006}
calculated photometric redshifts for $4.5 \mu$m-selected
sources in the ISS.  Clusters were then identified using a wavelet algorithm
which identified three-dimensional overdensities of galaxies (in 2 angular dimensions) 
with comparable photometric redshifts. The clusters have average halo
masses of $\sim 10^{14}$ $M_{\sun}$ \citep{Brodwin2007}.  The cluster
centers were determined from their peaks in the wavelet detection
map. Since this map is based on the positions of a limited number
of galaxies per cluster, the cluster center accuracy is expected
to be comparable to the $12\arcsec$ resolution of the wavelet map.

An estimate of the cluster redshifts was obtained from the photometric
redshift probability distribution of candidate cluster members.  At
$z < 0.5$, $67$ of the $91$ clusters candidates are confirmed by
the AGN and Galaxy Evolution Survey (AGES; Kochanek et al., in
prep.), which provides spectroscopic redshifts for $\sim 17000$
objects in the ISS. 
To date, $12$ of the $z > 1$ cluster candidates
have been confirmed spectroscopically \citep{Stanford2005, Elston2006,
Brodwin2006, Eisenhardt2008}.  Where available, we use the
spectroscopically determined cluster redshifts.  For the remaining
cluster sample, we use the best fit photometric redshift which
includes a small, empirically-determined offset relative to the
mean photometric redshift of the candidate cluster members.
This large, uniformly selected
galaxy cluster sample is spread over a large range of cosmic time
and thus provides a powerful tool for studying the dependence of
the AGN distribution on both redshift and environment.

\subsection{Mid-IR selected AGN}

Using mid-infrared colors, provided by the ISS, of nearly $10,000$
spectroscopically identified sources from AGES, \citet{Stern2005}
show that AGN can robustly be identified from their IRAC colors (see
also Lacy et al. 2004; Donley et al. 2008)\nocite{Donley2008}. Simple mid-infrared
color criteria select over $90$\% of the spectroscopically identified
type 1 AGN and $40$\% of the type 2 AGN. The selection relies on the fact 
that while stellar populations fade longward of their $1.6 \mu$m 
peak, luminous AGN exhibit red power-law emission throughout 
the mid-infrared. Notably, this power-law long wavelength emission 
is relatively immune to gas and dust absorption, making mid-infrared 
AGN selection sensitive to both obscured and unobscured luminous AGN, 
relatively independent of redshift. We note however that this selection
fails for low-luminosity AGN where the host galaxies contributes much of the 
mid-IR flux (Eckart et al., in prep.). We extract all 2262 sources selected by the Stern \etal
(2005) criteria that are well detected ($5\sigma$) in all four IRAC
bands.  For a typical AGN spectral energy distribution, the $5.8
\mu$m band is the least sensitive, with a $5 \sigma$ limiting depth
of $51 \mu$Jy ($3\arcsec$ diameter aperture). We find a mean surface
density of $283$ mid-infrared selected AGN candidates per square
degree in the Bo\"otes field.

\subsection{Radio selected AGN}

Several shallow ($\geq 1$mJy) radio surveys have covered the ISS
field at $1.4$ GHz, including the NRAO VLA Sky Survey (NVSS; Condon
\etal 1998\nocite{Condon1998}) and FIRST. More recently, the
Westerbork Synthesis Radio Telescope (WSRT) has covered approximately
$6.68$ square degrees of Bo\"{o}tes, reaching a $1 \sigma$ limit
of $28 \mu$Jy at $1.4$ GHz \citep{DeVries2002}. The final catalog
of $5 \sigma$ detected sources contains $3172$ objects\footnote[1]{The
catalog as well as the mosaic can be obtained through anonymous ftp
to ftp://ftp.nfra.nl/pub/Bootes.}, corresponding to a surface density
of $453$ radio sources per square degree.

The brightest radio sources in the sky tend to be AGN, with their
radio emission dominated by synchrotron radiation. However,
sufficiently deep radio surveys are also able to detect radio
emission associated with stellar processes such as supernova
explosions and compact stellar remnants, abundant in star-forming
galaxies. \citet{Condon1992} shows that for these star forming
galaxies, the star formation rate (SFR) is proportional to the radio
emission. Following Condon \etal (1992), Serjeant, Gruppioni \&
Oliver (2002)\nocite{Serjeant2002} obtain  $L_{\rm 1.4GHz}$(erg
s$^{-1}$ Hz$^{-1}$) $= 8.2$ x $10^{27}$ x SFR($M_{\sun}$ yr$^{-1}$).
Submillimeter galaxies correspond to the objects with the highest
known SFR, reaching SFR $\sim 1000 M_{\sun}$ yr$^{-1}$ \citep{Pope2006}.
We conservatively assume that no galaxy has a SFR greater than $3000
M_{\sun}$ yr$^{-1}$ and thus any radio source with a 1.4~GHz luminosity
density greater than $2.46$ x $10^{31}$ erg s$^{-1}$ Hz$^{-1}$ must
host an AGN (see \S 3). Given the depth of the WSRT observations, all radio-loud
AGN in the Bo\"{o}tes field should be detected out to redshift $z
\sim 3.7$.

\subsection{X-ray selected AGN}

XBo\"{o}tes is the largest contiguous survey with the {\it Chandra
X-Ray Observatory}. It consists of $9$ square degrees of imaging
of the NDWFS Bo\"{o}tes field, comprised of $126$ $5$~ks pointings,
with a corresponding depth of $\sim 7.8 \times 10^{-15}$ ergs
cm$^{-2}$ s$^{-1}$ in the $0.5$-$7$ keV band. The X-ray catalog
consists of $3293$ point sources with four or more X-ray counts
\citep[$366$ sources per square degree;][]{Kenter2005, Brand2006}
and is publicly
available\footnote[2]{http://heasarc.nasa.gov/W3Browse/all/xbootes.html}.
At the shallow depth of these data, the majority of the X-ray sources
are expected to be AGN.  Indeed, based on AGES optical spectra for
$892$ XBo\"{o}tes sources \citep{Kenter2005}, AGN represent at least
$69$\% of the sources with spectra.  Only $3$\% of the sources
are identified as stars.  The remaining $28$\% lack obvious AGN
features in their optical spectra but are suspected to be AGN based
on their X-ray luminosities. XBo\"{o}tes is only sensitive
to starbursts out to $z = 0.12$ \citep[e.g., see][]{Gorjian2008}.
Since only one of the galaxy clusters in the ISS sample is at $z
\le 0.12$, we consider the possible influence of X-ray detected
starbursts negligible in this work.

\section{Radial Surface Density Profile of AGN}

In order to study the cosmic evolution of AGN in galaxy clusters,
we opt for a simple empirical approach:  we calculate the radial
distribution of AGN around galaxy clusters as a function of cluster
redshift, with AGN identified using the three selection criteria
described above.  We separate the ISS clusters into three redshift
bins: $0 < z \le 0.5$, $0.5 < z \le 1$ and $1 < z \le 1.5$, resulting in
$91$, $140$ and $79$ clusters per bin, respectively, for the 
mid-IR and X-Ray techniques.  The remaining
$25$ clusters are at $z > 1.5$, and are thus ignored in this work.
The area covered by the WSRT radio data is smaller than the ISS and
thus only $285$ clusters have deep coverage at radio wavelengths.
Of these $285$ clusters, $77$, $121$ and $69$ are found at $0 < z
\le 0.5$, $0.5 < z \le 1$ and $1 < z \le 1.5$, respectively.

\begin{figure}[!t] 
\begin{center} 
\includegraphics[width=13cm]{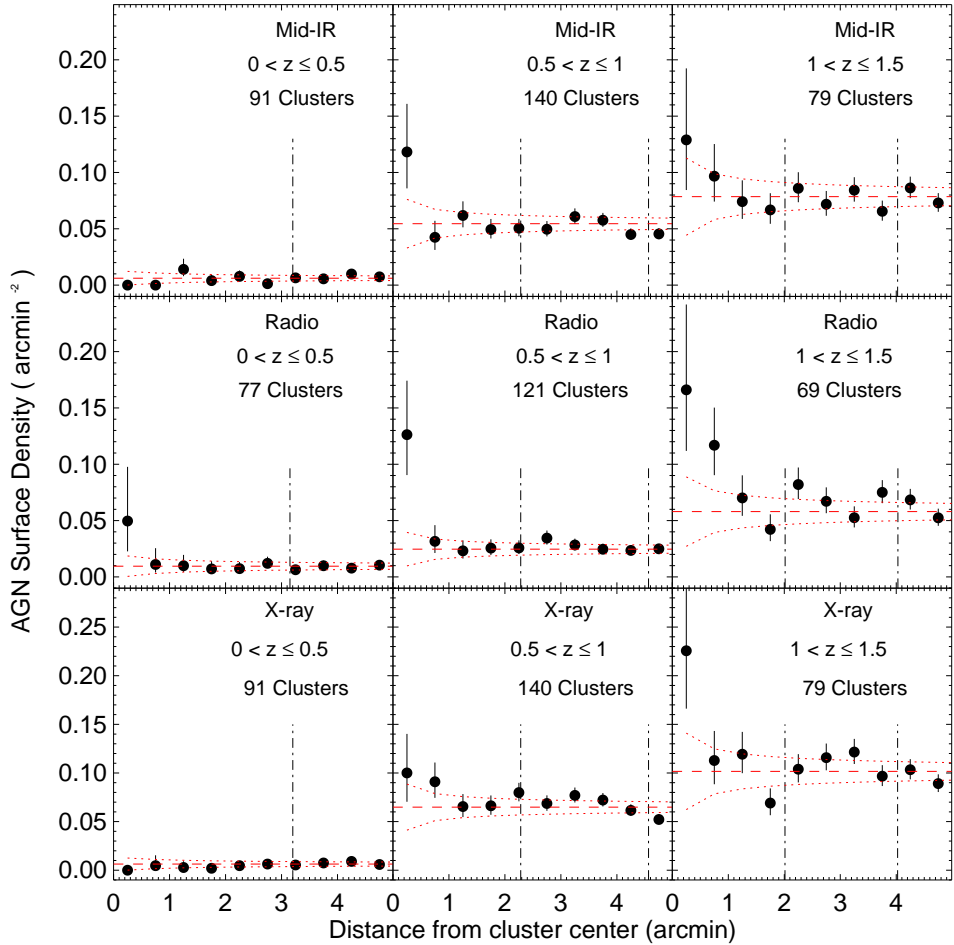}
\end{center}
\caption{Radial density profile of mid-IR selected AGN (top row), radio-selected AGN 
(middle row) and X-ray selected AGN (bottom row). Galaxy clusters are divided into three
redshift bins: $z \leq 0.5$ (left column), $0.5 < z \leq 1$ (middle column) and $1 < z \leq 1.5$ (right
column). The global field density is shown by the red dashed line. The red dotted
lines indicate the $1 \sigma$ variations in the field densities and the error bars indicate the $1 \sigma$
errors on the AGN counts; both assume Poissonian errors and Gehrels (1986) small numbers 
approximation. The $1$ Mpc and $2$ Mpc projected radii are indicated by vertical dot-dashed 
lines for the median redshift of the clusters plotted in each bin. The number of galaxy clusters 
used in the profile calculation is also indicated at the top right of each panel. An overdensity 
of AGN is observed in the centers of galaxy clusters at $z > 0.5$ while little or no variation relative 
to the field density is found at lower redshifts ($z \leq 0.5$).}
\label{plot}
\end{figure}

In order to ensure that similar populations are plotted in each
cluster redshift bin, we apply uniform luminosity cuts to the data.
In detail, since most of the sources considered here lack spectroscopic
redshifts, for a given cluster at redshift $z$, we only consider
sources brighter than some evolving flux density such that their
luminosity would be equal to or above our luminosity limit assuming
that the source were in the cluster. For the very sensitive radio
survey, we apply the luminosity cut $S_{\rm 1.4 GHz}>2.5 \times 
10^{40}$ ergs s$^{-1}$ (see \S 2.3) to isolate AGN from star-forming 
galaxies.  For the mid-IR and X-ray selected AGN, we apply a 
luminosity cut corresponding to the flux limit of
the corresponding surveys at $z = 1$ (see \S 3.1 and 3.3 for details).
This redshift is selected as a compromise between obtaining sufficient
numbers of AGN, large numbers of clusters, and sufficient leverage
on the cosmic time probed.  Table~1 presents the number of sources
that remain for each selection technique and redshift bin.  These
numbers are based on cuts (described in detail in the following
subsections) corresponding to the median redshift of the clusters
in each bin, e.g., $\langle z \rangle = 0.38$ ($0.39$), $\langle z \rangle = 0.74$ ($0.74$) and $\langle z \rangle
= 1.19$ ($1.18$) for mid-IR and X-ray (radio) selections.  Also
tabulated are the field AGN surface densities for each technique
and redshift bin, where we have simply divided the number of sources
by the corresponding survey area.  These field surface densities
are shown as dashed horizontal lines in Fig.~\ref{plot}.  Finally,
we note that for the X-ray and mid-IR selected AGN in the highest
redshift bin ($1 < z \leq 1.5$), this approach only provides a lower
limit to the surface density of AGN down to the luminosity limit
considered in the lower redshift bins.

The surface density distributions for AGN in the galaxy clusters
are calculated as follows.  We calculate the angular separation
from each ISS cluster center to each mid-infrared, radio and X-ray
selected AGN candidate, subject to the flux cuts discussed above.
AGN candidates are identified out to $5\arcmin$ from the cluster
centers, and we bin the radial distances to compute the surface
density profiles of AGN (0.5\arcmin\, bins), presented in Fig.~\ref{plot}.  As expected,
at large radii the derived surface densities asymptote to the field
AGN surface densities.  Throughout this paper, we adopt Poissonian
errors for AGN counts and use the \citet{Gehrels1986} small numbers
approximation for Poisson distributions.  Fig.~\ref{plot} presents
the $1 \sigma$ errors for both the AGN counts and the field surface
densities for each technique and redshift bin.

As a check on this method, we also calculated the AGN surface density
distribution at random locations in the field.  We created 40 mock
cluster samples by simply offsetting all of the cluster positions
by medium-scale offsets --- larger than the typical cluster size,
but small enough that few clusters would fall outside of the survey
region.  Deriving the radial surface density profile around the
mock cluster samples, we consistently find no significant variation
relative to the field surface densities.

In contrast to the mock sample, an excess of AGN is found near the
centers of clusters, most prominently at $z \geq 0.5$.  The next
sections detail the results obtained for each of the three AGN
selection criteria considered.  We note that we have verified that
spatially proximate clusters are not biasing our results. Since the
overdensities described below are only found at small radii ($< 1$
Mpc), the number of AGN counted as potential central members to
more than one cluster is negligible (null for $r < 0.5$ Mpc, and
$< 1\%$ for $0.5 < r < 1$ Mpc).

\subsection{Density profile of mid-IR selected AGN}

The radial surface density profile for mid-IR selected AGN is shown
in Fig.~\ref{plot} (top row), separated into the three redshift
bins described above.  The $5.8 \mu$m limiting flux density of the
ISS is $51 \mu$Jy ($5 \sigma$).  Assuming a pure power law spectrum
for the AGN, the corresponding luminosity density is $L_\nu = 4 \pi
d_L^2 S_\nu / (1+z)^{(1-\alpha)}$ where $d_L$ is the luminosity
distance and $\alpha$ is the spectral index ($S_\nu \propto
\nu^{\alpha}$).  This assumption of a power-law is born out by both
the mid-IR color-color plots of broad-lined AGN used to develop the
mid-IR AGN selection criteria (e.g., Lacy \etal 2004, Stern \etal
2005) and broad-band composite AGN spectral energy distributions
\citep{Richards2006}.  We adopt a mean spectral index of $0.73$,
as given by \citet{Stern2005}. The depth of the ISS $5.8 \mu$m data
corresponds to a $5.8 \mu$m luminosity of $2.2$ x $10^{44}$ ergs
s$^{-1}$ at $z = 1$, which we adopt as our luminosity cut.  Thus
for any given cluster we only consider mid-IR selected AGN whose luminosity (and
corresponding flux density) is above this value under the assumption
that they are at the cluster redshift.

The $z \leq 0.5$ galaxy clusters do not show any significant enhancement
of mid-IR selected AGN relative to the field.  In the $0.5 < z \leq 1$
bin, however, a weak overdensity of mid-IR selected AGN is observed
at the cluster center ($r < 0.3$ Mpc). Throughout the paper, the
significance of overdensities have been derived as follows.  We
first subtract the number of AGN in a given aperture by the field
AGN population to a given (redshift-dependent) flux density limit.
This field-corrected number of AGN is then referenced to the expected
field counts using the \citet{Gehrels1986} small numbers approximation
for Poisson distributions.  The central excess of $0.5 < z \leq 1$
mid-IR selected AGN is thus significant at the $2 \sigma$ level.
A small excess of AGN is suggested in the center of the $z > 1$
galaxy clusters, but is not significant relative to the field
density.  Recall that since the highest redshift bin considers more
luminous AGN than the $z < 1$ bins, the resulting, weak overdensity
is a lower limit to the number of AGN per cluster to the luminosity
limit of the $z < 1$ clusters.

\subsection{Density profile of radio selected AGN}

The radial surface density profile for radio-selected AGN is shown
in Fig.~\ref{plot} (middle row), separated into three redshift bins.
For each redshift bin, we apply the conservative uniform luminosity
cut described in \S 2.3 to remove star-forming galaxies, adopting
a mean radio spectral index of $-0.75$, as given in de Vries \etal (2002).

In the $z \leq 0.5$ bin, a weak overdensity ($1.5 \sigma$) of radio
sources is observed at the cluster center. The $0.5 < z \leq 1$ galaxy
clusters show a very clear excess of AGN in the center ($r < 0.3$
Mpc), with no excess seen at larger radii.  The central excess is
significant at the $3 \sigma$ level. The higher redshift galaxy
clusters ($1 < z \leq 1.5$) also show a significant excess ($2 \sigma$)
at the cluster center, albeit weaker and on a larger scale ($r <
0.5$ Mpc) relative to the moderate redshift galaxy clusters.

\subsection{Density profile of X-ray selected AGN}

The radial surface density profile for X-ray selected AGN is shown
in Fig.~\ref{plot} (bottom row).  The limiting flux of XBo\"{o}tes
in the $0.5-7.0$ keV range is $\sim 7.8$ x $10^{-15}$ ergs cm$^{-2}$
s$^{-1}$, where the XBo\"{o}tes flux values have been calculated
from the counts assuming an X-ray AGN power-law spectrum with photon
index $\Gamma = 1.7$ (Kenter \etal 2005).  While the shallow depth
of the XBo\"{o}tes survey implies that any source detected at $z
\geq 0.12$ must be an AGN \citep{Gorjian2008}, we only plot sources
whose X-ray luminosity exceeds the limiting sensitivity of XBo\"{o}tes
at redshift unity. The same photon index value is adopted to calculate
the X-ray luminosities. The luminosity cut applied for X-ray selection
is therefore $3.3$ x $10^{43}$ ergs s$^{-1}$ in the $0.5-7.0$ keV
range.  As emphasized for the mid-IR selected AGN, the lower, right
panel of Fig.~\ref{plot} is a lower limit to the surface density
of X-ray selected AGN at the luminosities plotted in the first two
panels.

No significant enhancement of the X-ray selected AGN is observed
for the lowest redshift ($z \leq 0.5$) galaxy clusters. A small central
overdensity is suggested for $0.5 < z \leq 1$ galaxy clusters ($1.2
\sigma$ level; $r < 0.5$ Mpc). On the other hand, the $1 < z \leq
1.5$ galaxy clusters show a modest excess ($2 \sigma$) at the cluster
center ($r < 0.25$ Mpc).


We note that, in principle, X-ray point sources are harder to
identify in galaxy clusters with significant diffuse X-ray emission.
The intergalactic medium in such clusters would tend to decrease
the apparent AGN surface density for lower redshift clusters in a
shallow survey such as XBo\"{o}tes. XBo\"{o}tes identifies $43$
extended sources \citep{Kenter2005} of which $38$ are covered by
ISS.  The size of these extended sources, derived from Gaussian
fits to their source profiles, is (except for one source) less than
$0.5'$.  The overdensity of X-ray point sources that we observe in
our galaxy cluster sample is clearly detected within $0.5$ Mpc ($r
\sim 1.5'$ at $z = 0.5$) of the cluster center.  The brightness of
an unresolved cluster member with an X-ray luminosity equivalent
to our threshold would, on average, be five times brighter than
that of the extended sources.  Only in four clusters would a threshold
source have a flux comparable to the extended source (e.g., brightness
contrast $< 2$).  Given these small numbers, we are confident that
extended, diffuse X-ray emission from rich clusters is not strongly
biasing our results.


\section{Discussion}

This paper presents the first comprehensive study of the radial
density profile of AGN for a uniformly-selected cluster sample
distributed over a large fraction of cosmic time.  Only the
radio-selected AGN show a weak ($1.5 \sigma$) overdensity in the
lowest redshift ($z \leq 0.5$) bin.  An excess of AGN is
observed near the center ($r < 0.5$ Mpc) of galaxy clusters at $z
> 0.5$, and all three AGN selection criteria show a pronounced enhancement
in AGN activity at $0.5 < z \leq 1$ relative to $z \leq 0.5$.
The excess is even observed for the highest redshift galaxy clusters
($1 < z \leq 1.5$), though we are only plotting to the flux limits
of the input mid-IR and X-ray catalogs at these high redshifts,
implying a less sensitive luminosity limit than in the $z \leq 1$
clusters.  We therefore show that AGN activity in galaxy
clusters increases with redshift.


We first compare the luminosity cuts applied for each selection
criteria to see how the AGN samples relate to one another.  The
luminosity cut applied for the mid-infrared selection is $2.2$ x
$10^{44}$ ergs s$^{-1}$ at $5.8 \mu$m. \citet{Elvis1994} provides
a mean spectral energy distribution (SED) for quasars derived from
broad-band photometry covering the full electromagnetic spectrum
from radio to X-rays. Using this composite SED for radio-loud
quasars, the luminosity cut at $5.8 \mu$m corresponds to a luminosity
cut of $\sim 2.2 \times 10^{43}$ ergs s$^{-1}$ in the $0.5-7$ keV X-ray
band and $2.2 \times 10^{41}$ ergs s$^{-1}$ at $1.4$ GHz. Our X-ray
luminosity cut is $3.3 \times 10^{43}$ ergs s$^{-1}$ which is
comparable to the depth of the mid-infrared selection. A more recent
determination of the mean quasar SED was performed by \citet{Richards2006}
and gives an X-ray luminosity five times smaller at a given
mid-infrared luminosity compared to the Elvis \etal (1994) template.
This would imply that our X-ray selection only identifies AGN five times
more powerful than the mid-infrared selection.
Our radio luminosity cut is $2.5 \times 10^{40}$ ergs s$^{-1}$ at $1.4$ GHz.  Therefore,
the AGN radio selection depth is reaching almost ten times the depth
of the mid-infrared selection, and ten to fifty times the depth of
the X-ray selection, depending on which quasar composite template
is adopted.  This deeper radio sensitivity is a plausible explanation
for why only the radio-selected AGN show an overdensity in the
lowest redshift galaxy clusters.

One concern is that due to the larger volumes accessible
at higher redshifts, our sample of clusters becomes progressively
richer with redshift.  In this case, the rising number of AGN seen
in the higher redshift clusters could simply trace the rising number
of massive cluster members in the higher redshift clusters, with a
constant, non-evolving fraction of massive galaxies being active.
In order to study this effect, we use the photometric redshifts
derived from the 4.5$\mu$m ISS sample (see \S 2.1) to determine the
mean number of cluster members brighter than $0.5 L^*$ as a function
of cluster redshift.  We consider the observed $3.6 \mu$m flux
density, converted to $1.6 \mu$m rest-frame luminosity assuming a
$z_f = 3$ formation redshift and passive evolution \citep[e.g., see
][]{Eisenhardt2008}.  We only consider members brighter than $0.5 L^*$ within the inner
$0.5$ Mpc of the clusters.  As expected, we do indeed find clusters
at $z \sim 0.3$ are, on average, slightly less rich than clusters
at $z \sim 0.8$.  Clusters at $0 < z \leq 0.5$ have an average of
$6.5$ members brighter than $0.5 L^*$ while $10.0$ members are found,
on average, in clusters at $0.5 < z \leq 1$.  At redshifts above 0.8,
the number of cluster members in the inner $0.5$ Mpc and brighter
than $0.5 L^*$ actually begins to decrease, with $5.7$ galaxies,
on average, for clusters at $z > 1$.  This is not a sensitivity
issue since the $90$ seconds of combined exposure of ISS provides
sufficient sensitivity to detect evolving $L^*$ galaxies to $z =
2$ \citep{Eisenhardt2008}.  This decrease in cluster member numbers instead
could reflect that clusters at $z > 0.8$ are still in the process
of collapsing, that the clusters members are merging, or some combination 
thereof. However, the variation with redshift is only at the 50\% level, while the
cosmic evolution of AGN activity is a factor of several more
significant (with some variation depending on which AGN selection
criterion has been used).  For the following discussion, we therefore
omit the modest systematic dependence of cluster richness on redshift
inherent to the Bo\"{o}tes cluster sample.  We do note, however,
that this trend would serve to make the fraction of active galaxies
in the central regions of $z > 1$ clusters even higher relative to
the fraction at $z < 0.8$, since the number of AGN is rising while
the number of luminous cluster members is falling.

\begin{figure}[!t] 
\begin{center} 
\includegraphics[width=13cm]{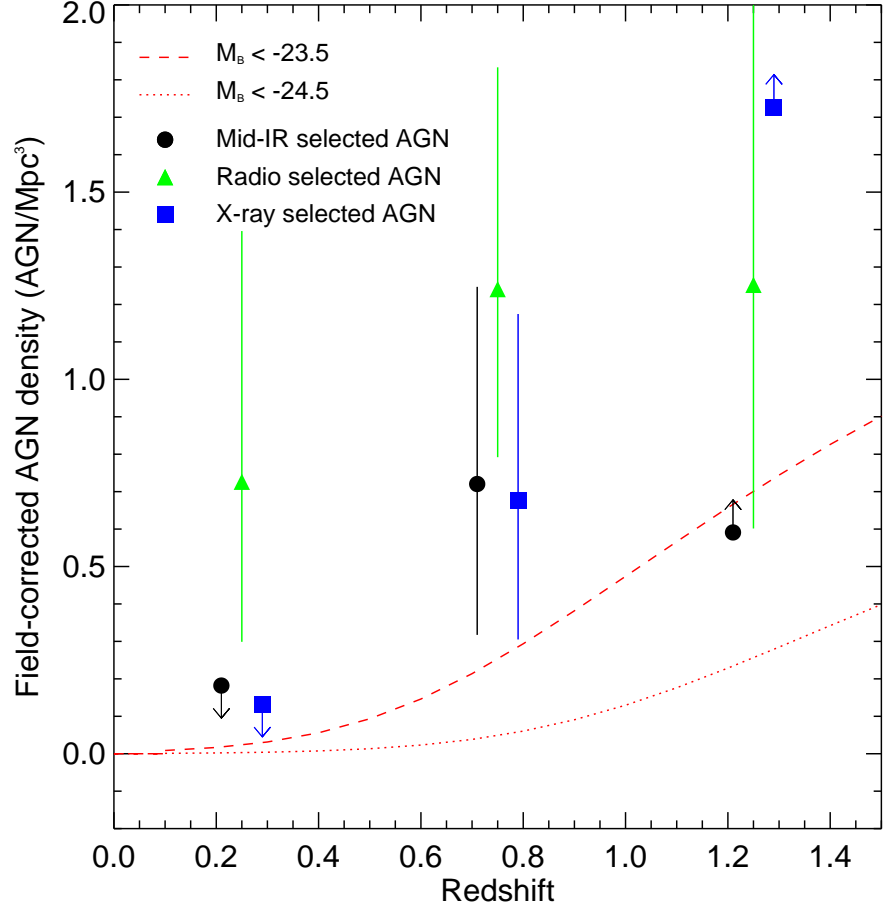}
\end{center}
\caption{Field-corrected volume density of AGN in galaxy clusters and comparison of its 
evolution relative to an optical quasar luminosity function. Points show the field-corrected overdensity of 
AGN within $0.25$ Mpc of the ISS cluster centers for the three AGN criteria: 
mid-infrared (black points), radio (green triangles) and X-ray (blue squares). 
No mid-IR or X-ray selected AGN are found within $0.25$ Mpc for the galaxy 
clusters at $z < 0.5$. We plot the 1$\sigma$ uncertainty of the field density 
as an upper limit. Lines show the relative evolution of quasars brighter than $L_{\rm AGN}^*$ 
($M_B = -24.5$, dotted red line) and $L_{\rm AGN}^* + 1$ ($M_B = -23.5$, dashed red line) 
for the field, where $L_{\rm AGN}^*$ is calculated at $z=1$. These models, which show a less dramatic 
evolution of AGN in the field relative to clusters, are calculated from the Croom et al. (2004) 2dFQZ 
QLF (see text for details).}
\label{plot2}
\end{figure}

It is well known that AGN are more common at high redshift. Indeed,
there is a strong increase in the quasar density with redshift from
the local universe out to $z \sim 2$, a result that dates back four
decades \citep{Schmidt1968}. Since then, numerous studies have
determined the quasar luminosity function (QLF) for radio (e.g.,~Dunlop
\& Peacock 1990\nocite{Dunlop1990}), optical \citep[e.g.,][]{Croom2004},
infrared \citep[e.g.,][]{Brown2006} and X-ray AGN samples
\citep[e.g.,][]{Ueda2003}.  \citet{Eastman2007} find that the cluster
AGN fraction at $z \sim 0.6$ is approximately a factor of 20 greater
than the cluster AGN fraction at $z \sim 0.2$. In comparison, this
increase is only $1.5$ ($3.3$) for the field sample of AGN more
luminous than $L_x > 10^{42}$ ($10^{43}$) ergs s$^{-1}$ over this
redshift range. In order to compare the AGN overdensities associated
with galaxy clusters, seen in Fig.~\ref{plot}, to the field AGN
population, we derive average cluster AGN volume densities as
follows.  We count the number of sources projected within $0.25$~Mpc
of the cluster centers for the various redshift bins and selection
criteria, and then field-correct this number by subtracting the
expected number of sources over this same area based on the field
surface density.  We assume any identified overdensity is associated
with the cluster (e.g., is neither foreground nor background).  For
the sake of simplicity, we assume a spherical distribution at the
cluster redshift to derive a volume density from the measured
(field-corrected) surface density.  Fig.~\ref{plot2} presents this
field-corrected volume density of AGN within $0.25$~Mpc of the
galaxy clusters in the three redshift intervals and for the three
AGN selection criteria.  We also plot the evolution of optically-selected
quasars over this same redshift range using the QLF derived by
\citet{Croom2004} from the 2dFQZ project\footnote[3]{Croom et al.
(2004) present various forms of the empirical QLF.  We use the
version which fits the QLF with a second-order polynomial to describe
the luminosity evolution (Table~5 of that paper).}. At $z=1$,
\citet{Croom2004} find that $L_{\rm AGN}^* = 1.6 \times 10^{45}\,
{\rm ergs}\, {\rm s}^{-1}$, corresponding to $M_B = -24.5$.  Using
the \citet{Elvis1994} quasar template, the mid-IR and X-ray depth
of our survey correspond to optical luminosities of $L_{\rm AGN}^*
+ 1$ ($M_B = -23.5$).  Therefore, we plot in Fig.~\ref{plot2} the
cosmic evolution of optically-selected field quasars brighter than
$L_{\rm AGN}^* + 1$ ($M_B = -23.5$) and $L_{\rm AGN}^*$ ($M_B =
-24.5$).  We find a more pronounced cosmic evolution in the relative
density of cluster AGN compared to the field from $z = 0$ to $1.5$.
In particular, the field-corrected X-ray selected AGN density in
clusters evolves at least three times more rapidly than that of the
field population.

This work shows that a higher surface density of AGN is observed
in galaxy cluster centers than in the field and that this AGN
overdensity increases with redshift out to $z \ge 1$. We therefore
confirm a Butcher-Oemler type effect for AGN in galaxy clusters
\citep[e.g.,][]{Martini2007} as the number of AGN in clusters
increases with redshift similar to the way that higher redshift
galaxy clusters have larger numbers of blue, star-forming galaxies.

We find that up to ten percent of the galaxy clusters have at least
one AGN near their center ($r < 0.25$ Mpc). Table~2 summarizes the
percentage of galaxy clusters with an AGN for each selection technique
and redshift bin. These fractions are field-corrected.  Considering
the luminosity cuts applied for each selection, only $3$\% of galaxy
clusters at $z < 0.5$ have a radio-selected AGN near their center
($r< 0.25$ Mpc) and neither mid-IR nor X-ray selected AGN are found.
We find that the fraction of galaxy clusters hosting such an AGN
evolves with redshift.  By $z \sim 1.5$, up to $10$\% of our clusters
sample host at least one X-ray emitting AGN near their center ($9$\%
for radio-selected AGN).

 
Finally, we note that the current analysis is rather simple, using only the apparent 
two dimensional distribution of sources across the sky to address the evolving 
role of AGN in clusters.  In principle, a three dimensional analysis, accounting 
for source redshifts (either spectroscopic or photometric), would significantly 
lower the field counts in each redshift bin and provide a more robust 
measurement of the AGN Butcher-Oemler effect.  We have opted for the former 
analysis as it relies only on the simple observables of position and flux, and is 
thus model free. Brodwin \etal (2006) derive photometric redshifts for the Bo\"{o}tes field 
AGN with an accuracy $\sigma = 0.12 (1 + z)$, with a $5$\% outlier fraction. This is
approximately half the precision that can be derived for normal galaxies. We are currently
obtaining spectroscopic observations of candidate active members of clusters. Future work 
will use these results to better constrain the pronounced cosmic evolution of AGN activity in 
rich environments.

\section{Summary}

The spatial distribution of AGN has been derived for a well-defined
sample of galaxy clusters at $0 < z \leq 1.5$. The AGN are selected
using three different selection techniques (mid-IR color, radio
luminosity, and X-ray luminosity) that isolate distinct types of
active galaxies to ensure the fullest selection of AGN in our galaxy
cluster sample. An overdensity of AGN is found near the center ($r
< 0.5$ Mpc) of galaxy clusters at $z > 0.5$ and this AGN excess
increases with redshift. We confirm that the rising number of
AGN is not simply due to the clusters becoming richer with redshift
with a non-evolving fraction of cluster members being active. It
is well known that AGN are also more common at high redshift. We
therefore also confirm that the observed increase in redshift is
more pronounced in the galaxy cluster sample than in the field. We
thus find a Butcher-Oemler type effect for AGN in galaxy clusters.


We thank the anonymous referee whose comments significantly improved
this paper. We thank Joel Vernet and Carlos De Breuck for carefully
reading the manuscript. This work is partly based on observations
made with the {\em Spitzer Space Telescope}, which is operated by
the Jet Propulsion Laboratory, California Institute of Technology
under a contract with NASA. This work made use of data products
provided by the NOAO Deep Wide-Field Survey, which is supported by
the National Optical Astronomy Observatory (NOAO). NOAO is operated
by AURA, Inc., under a cooperative agreement with the National
Science Foundation. AHG acknowledges support from NSF grant AST-0708490. 
SAS's work was performed under the auspices of the U.S. Department of Energy, 
National Nuclear Security Administration by the University of California, 
Lawrence Livermore National
Laboratory under contract No. W-7405-Eng-48.

\newpage

\bibliographystyle{apj}
\bibliography{apj-jour,/Users/audreygalametz/biblio/biblio}

\begin{thebibliography}{58}
\expandafter\ifx\csname natexlab\endcsname\relax\def\natexlab#1{#1}\fi

\bibitem[{{Bahcall} {et~al.}(1969){Bahcall}, {Schmidt}, \&
  {Gunn}}]{Bahcall1969}
{Bahcall}, J.~N., {Schmidt}, M., \& {Gunn}, J.~E. 1969, \apjl, 157, L77

\bibitem[{{Becker} {et~al.}(1995){Becker}, {White}, \& {Helfand}}]{Becker1995}
{Becker}, R.~H., {White}, R.~L., \& {Helfand}, D.~J. 1995, \apj, 450, 559

\bibitem[{{Best} {et~al.}(2005){Best}, {Kauffmann}, {Heckman}, {Brinchmann},
  {Charlot}, {Ivezi{\'c}}, \& {White}}]{Best2005}
{Best}, P.~N., {Kauffmann}, G., {Heckman}, T.~M., {Brinchmann}, J., {Charlot},
  S., {Ivezi{\'c}}, {\v Z}., \& {White}, S.~D.~M. 2005, \mnras, 362, 25

\bibitem[{{Bignamini} {et~al.}(2008){Bignamini}, {Tozzi}, {Borgani}, {Ettori},
  \& {Rosati}}]{Bignamini2008}
{Bignamini}, A., {Tozzi}, P., {Borgani}, S., {Ettori}, S., \& {Rosati}, P.
  2008, \aap, 807

\bibitem[{{B{\^i}rzan} {et~al.}(2004){B{\^i}rzan}, {Rafferty}, {McNamara},
  {Wise}, \& {Nulsen}}]{Birzan2004}
{B{\^i}rzan}, L., {Rafferty}, D.~A., {McNamara}, B.~R., {Wise}, M.~W., \&
  {Nulsen}, P.~E.~J. 2004, \apj, 607, 800

\bibitem[{{Brand} {et~al.}(2006)}]{Brand2006}
{Brand}, K. {et~al.} 2006, \apj, 641, 140

\bibitem[{{Brodwin} {et~al.}(2007){Brodwin}, {Gonzalez}, {Moustakas},
  {Eisenhardt}, {Stanford}, {Stern}, \& {Brown}}]{Brodwin2007}
{Brodwin}, M., {Gonzalez}, A.~H., {Moustakas}, L.~A., {Eisenhardt}, P.~R.,
  {Stanford}, S.~A., {Stern}, D., \& {Brown}, M.~J.~I. 2007, \apjl, 671, L93

\bibitem[{{Brodwin} {et~al.}(2006)}]{Brodwin2006}
{Brodwin}, M. {et~al.} 2006, \apj, 651, 791

\bibitem[{{Brodwin} {et~al.}(2008)}]{Brodwin2008}
---. 2008, astro-ph: arXiv0810.0528

\bibitem[{{Brown} {et~al.}(2006)}]{Brown2006}
{Brown}, M.~J.~I. {et~al.} 2006, \apj, 638, 88

\bibitem[{{Cappi} {et~al.}(2001)}]{Cappi2001}
{Cappi}, M. {et~al.} 2001, \apj, 548, 624

\bibitem[{{Coil} {et~al.}(2007){Coil}, {Hennawi}, {Newman}, {Cooper}, \&
  {Davis}}]{Coil2007}
{Coil}, A.~L., {Hennawi}, J.~F., {Newman}, J.~A., {Cooper}, M.~C., \& {Davis},
  M. 2007, \apj, 654, 115

\bibitem[{{Condon}(1992)}]{Condon1992}
{Condon}, J.~J. 1992, \araa, 30, 575

\bibitem[{{Condon} {et~al.}(1998){Condon}, {Cotton}, {Greisen}, {Yin},
  {Perley}, {Taylor}, \& {Broderick}}]{Condon1998}
{Condon}, J.~J., {Cotton}, W.~D., {Greisen}, E.~W., {Yin}, Q.~F., {Perley},
  R.~A., {Taylor}, G.~B., \& {Broderick}, J.~J. 1998, \aj, 115, 1693

\bibitem[{{Croft} {et~al.}(2007){Croft}, {de Vries}, \& {Becker}}]{Croft2007}
{Croft}, S., {de Vries}, W., \& {Becker}, R.~H. 2007, \apjl, 667, L13

\bibitem[{{Croom} {et~al.}(2004){Croom}, {Smith}, {Boyle}, {Shanks}, {Miller},
  {Outram}, \& {Loaring}}]{Croom2004}
{Croom}, S.~M., {Smith}, R.~J., {Boyle}, B.~J., {Shanks}, T., {Miller}, L.,
  {Outram}, P.~J., \& {Loaring}, N.~S. 2004, \mnras, 349, 1397

\bibitem[{{Croom} {et~al.}(2005)}]{Croom2005}
{Croom}, S.~M. {et~al.} 2005, \mnras, 356, 415

\bibitem[{{de Vries} {et~al.}(2002){de Vries}, {Morganti}, {R{\"o}ttgering},
  {Vermeulen}, {van Breugel}, {Rengelink}, \& {Jarvis}}]{DeVries2002}
{de Vries}, W.~H., {Morganti}, R., {R{\"o}ttgering}, H.~J.~A., {Vermeulen}, R.,
  {van Breugel}, W., {Rengelink}, R., \& {Jarvis}, M.~J. 2002, \aj, 123, 1784

\bibitem[{{Dey} {et~al.}(2008)}]{Dey2008}
{Dey}, A. {et~al.} 2008, \apj, 677, 943

\bibitem[{{Dickinson}(1993)}]{Dickinson1993}
{Dickinson}, M. 1993, in Astronomical Society of the Pacific Conference Series,
  Vol.~51, Observational Cosmology, ed. G.~L. {Chincarini}, A.~{Iovino},
  T.~{Maccacaro}, \& D.~{Maccagni}, 434

\bibitem[{{Donley} {et~al.}(2008){Donley}, {Rieke}, {Perez-Gonzalez}, \&
  {Barro}}]{Donley2008}
{Donley}, J.~L., {Rieke}, G.~H., {Perez-Gonzalez}, P.~G., \& {Barro}, G. 2008,
  astro-ph: arXiv:0806.4610

\bibitem[{{Dressler} {et~al.}(1985){Dressler}, {Thompson}, \&
  {Shectman}}]{Dressler1985}
{Dressler}, A., {Thompson}, I.~B., \& {Shectman}, S.~A. 1985, \apj, 288, 481

\bibitem[{{Dunlop} \& {Peacock}(1990)}]{Dunlop1990}
{Dunlop}, J.~S. \& {Peacock}, J.~A. 1990, \mnras, 247, 19

\bibitem[{{Eastman} {et~al.}(2007){Eastman}, {Martini}, {Sivakoff}, {Kelson},
  {Mulchaey}, \& {Tran}}]{Eastman2007}
{Eastman}, J., {Martini}, P., {Sivakoff}, G., {Kelson}, D.~D., {Mulchaey},
  J.~S., \& {Tran}, K.-V. 2007, \apjl, 664, L9

\bibitem[{{Ebeling} {et~al.}(2001){Ebeling}, {Edge}, \& {Henry}}]{Ebeling2001}
{Ebeling}, H., {Edge}, A.~C., \& {Henry}, J.~P. 2001, \apj, 553, 668

\bibitem[{{Eckart} {et~al.}(2006){Eckart}, {Stern}, {Helfand}, {Harrison},
  {Mao}, \& {Yost}}]{Eckart2006}
{Eckart}, M.~E., {Stern}, D., {Helfand}, D.~J., {Harrison}, F.~A., {Mao},
  P.~H., \& {Yost}, S.~A. 2006, \apjs, 165, 19

\bibitem[{{Eisenhardt} {et~al.}(2004)}]{Eisenhardt2004}
{Eisenhardt}, P.~R. {et~al.} 2004, \apjs, 154, 48

\bibitem[{{Eisenhardt} {et~al.}(2008)}]{Eisenhardt2008}
{Eisenhardt}, P.~R.~M. {et~al.} 2008, \apj, 684, 905

\bibitem[{{Elston} {et~al.}(2006)}]{Elston2006}
{Elston}, R.~J. {et~al.} 2006, \apj, 639, 816

\bibitem[{{Elvis} {et~al.}(1994)}]{Elvis1994}
{Elvis}, M. {et~al.} 1994, \apjs, 95, 1

\bibitem[{{Gehrels}(1986)}]{Gehrels1986}
{Gehrels}, N. 1986, \apj, 303, 336

\bibitem[{{Gorjian} {et~al.}(2008)}]{Gorjian2008}
{Gorjian} {et~al.} 2008, \apj, 679

\bibitem[{{Gunn} \& {Weinberg}(1995)}]{Gunn1995}
{Gunn}, J. \& {Weinberg}, D. 1995, in Wide Field Spectroscopy and the Distant
  Universe, ed. S.~J. {Maddox} \& A.~{Aragon-Salamanca}, 3

\bibitem[{{Jannuzi} \& {Dey}(1999)}]{Januzzi1999}
{Jannuzi}, B.~T. \& {Dey}, A. 1999, in Astronomical Society of the Pacific
  Conference Series, Vol. 191, Photometric Redshifts and the Detection of High
  Redshift Galaxies, ed. R.~{Weymann}, L.~{Storrie-Lombardi}, M.~{Sawicki}, \&
  R.~{Brunner}, 111

\bibitem[{{Kenter} {et~al.}(2005)}]{Kenter2005}
{Kenter}, A. {et~al.} 2005, \apjs, 161, 9

\bibitem[{{Lacy} {et~al.}(2004)}]{Lacy2004}
{Lacy}, M. {et~al.} 2004, \apjs, 154, 166

\bibitem[{{Lin} \& {Mohr}(2007)}]{Lin2007}
{Lin}, Y.-T. \& {Mohr}, J.~J. 2007, \apjs, 170, 71

\bibitem[{{Magliocchetti} {et~al.}(2008)}]{Magliocchetti2008}
{Magliocchetti}, M. {et~al.} 2008, \mnras, 383, 1131

\bibitem[{{Magorrian} {et~al.}(1998)}]{Magorrian1998}
{Magorrian}, J. {et~al.} 1998, \aj, 115, 2285

\bibitem[{{Martini} {et~al.}(2006){Martini}, {Kelson}, {Kim}, {Mulchaey}, \&
  {Athey}}]{Martini2006}
{Martini}, P., {Kelson}, D.~D., {Kim}, E., {Mulchaey}, J.~S., \& {Athey}, A.~A.
  2006, \apj, 644, 116

\bibitem[{{Martini} {et~al.}(2002){Martini}, {Kelson}, {Mulchaey}, \&
  {Trager}}]{Martini2002}
{Martini}, P., {Kelson}, D.~D., {Mulchaey}, J.~S., \& {Trager}, S.~C. 2002,
  \apjl, 576, L109

\bibitem[{{Martini} {et~al.}(2007){Martini}, {Mulchaey}, \&
  {Kelson}}]{Martini2007}
{Martini}, P., {Mulchaey}, J.~S., \& {Kelson}, D.~D. 2007, \apj, 664, 761

\bibitem[{{Matthews} {et~al.}(1964){Matthews}, {Morgan}, \&
  {Schmidt}}]{Matthews1964}
{Matthews}, T.~A., {Morgan}, W.~W., \& {Schmidt}, M. 1964, \apj, 140, 35

\bibitem[{{McNamara} {et~al.}(2000)}]{McNamara2000}
{McNamara}, B.~R. {et~al.} 2000, \apjl, 534, L135

\bibitem[{{Miley} {et~al.}(2004)}]{Miley2004}
{Miley}, G.~K. {et~al.} 2004, \nat, 427, 47

\bibitem[{{Murray} {et~al.}(2005)}]{Murray2005}
{Murray}, S.~S. {et~al.} 2005, \apjs, 161, 1

\bibitem[{{Pope} {et~al.}(2006){Pope}, {Chary}, {Dickinson}, \&
  {Scott}}]{Pope2006}
{Pope}, A., {Chary}, R., {Dickinson}, M., \& {Scott}, D. 2006, in Bulletin of
  the American Astronomical Society, Vol.~38, Bulletin of the American
  Astronomical Society, 1172

\bibitem[{{Richards} {et~al.}(2006)}]{Richards2006}
{Richards}, G.~T. {et~al.} 2006, \apjs, 166, 470

\bibitem[{{Ruderman} \& {Ebeling}(2005)}]{Ruderman2005}
{Ruderman}, J.~T. \& {Ebeling}, H. 2005, \apjl, 623, L81

\bibitem[{{Schmidt}(1968)}]{Schmidt1968}
{Schmidt}, M. 1968, \apj, 151, 393

\bibitem[{{Serjeant} {et~al.}(2002){Serjeant}, {Gruppioni}, \&
  {Oliver}}]{Serjeant2002}
{Serjeant}, S., {Gruppioni}, C., \& {Oliver}, S. 2002, \mnras, 330, 621

\bibitem[{{Stanford} {et~al.}(2005)}]{Stanford2005}
{Stanford}, S.~A. {et~al.} 2005, \apjl, 634, L129

\bibitem[{{Stern} {et~al.}(2000){Stern}, {Djorgovski}, {Perley}, {de Carvalho},
  \& {Wall}}]{Stern2000}
{Stern}, D., {Djorgovski}, S.~G., {Perley}, R.~A., {de Carvalho}, R.~R., \&
  {Wall}, J.~V. 2000, \aj, 119, 1526

\bibitem[{{Stern} {et~al.}(2003){Stern}, {Holden}, {Stanford}, \&
  {Spinrad}}]{Stern2003}
{Stern}, D., {Holden}, B., {Stanford}, S.~A., \& {Spinrad}, H. 2003, \aj, 125,
  2759

\bibitem[{{Stern} {et~al.}(2005)}]{Stern2005}
{Stern}, D. {et~al.} 2005, \apj, 631, 163

\bibitem[{{Ueda} {et~al.}(2003){Ueda}, {Akiyama}, {Ohta}, \&
  {Miyaji}}]{Ueda2003}
{Ueda}, Y., {Akiyama}, M., {Ohta}, K., \& {Miyaji}, T. 2003, \apj, 598, 886

\bibitem[{{Venemans} {et~al.}(2005)}]{Venemans2005}
{Venemans}, B.~P. {et~al.} 2005, \aap, 431, 793

\bibitem[{{Venemans} {et~al.}(2007)}]{Venemans2007}
---. 2007, \aap, 461, 823

\end{thebibliography}

\newpage


\begin{deluxetable}{lcccccc}
\tablecaption{Number and surface density of Bo\"{o}tes AGN candidates after luminosity cuts.}
\tablecolumns{7}
\tablehead{
\colhead{} &
\colhead{Luminosity limit} &
\colhead{Area} &
\colhead{Total number} &
\colhead{ $0 < z \leq 0.5$} &
\colhead{$0.5 < z \leq 1$} &
\colhead{$1 < z \leq 1.5$} \\
\colhead{Selection} &
\colhead{(ergs s$^{-1}$)} &
\colhead{(deg$^2$)} &
\colhead{N ($\Sigma$, arcmin$^{-2}$)} &
\colhead{N$_{\rm AGN}$ ($\Sigma_{\rm AGN}$)} &
\colhead{N$_{\rm AGN}$ ($\Sigma_{\rm AGN}$)} &
\colhead{N$_{\rm AGN}$ ($\Sigma_{\rm AGN}$)}}
\startdata
Mid-IR	& $S_{\rm 5.8 \mu m}>2.2 \times 10^{44}$ & 8.0 &	2262	~(0.079)	&	177~(0.006)	&	1570~(0.054)	&	2262~(0.079)	\\
Radio	& $S_{\rm 1.4 GHz}>2.5 \times 10^{40}$   & 6.68 &	3172~(0.132)	&	229~(0.009) 	&	591~(0.025)	&	1394~(0.058)	\\ 
X-ray	& $S_{\rm 0.5-7 keV}>3.3 \times 10^{43}$ & 9.0 &	3293	~(0.102)	&	205~(0.006)	&	2102~(0.065)	&	3293~(0.102)	\\
\enddata
\label{table1}
\end{deluxetable}


\begin{deluxetable}{lccc}
\tablecaption{Fraction of galaxy clusters with at least one AGN in the inner $0.25$ Mpc}
\tablehead{
\colhead{Selection} &
\colhead{ $0 < z \leq 0.5$} &
\colhead{$0.5 < z \leq 1$} &
\colhead{$1 < z \leq 1.5$}}
\startdata
Mid-IR	&	$<1$\%			&  $5 \pm 2$\% 	&  $4 \pm 3$\%	\\
Radio	&	$3 \pm 2$\%		&  $8 \pm 3$\%		&  $9 \pm 4$\%	\\
X-ray	&	$<1$\%			&  $3 \pm 2$\%		&  $10 \pm 4$\%	\\	 
\enddata
\label{table2}
\end{deluxetable}

\newpage

\newpage

\end{document}